\documentclass[a4,11pt]{article}
\usepackage{amsfonts}

\newtheorem{lemma}{Lemma}
\newtheorem{theorem}{Theorem}%[section]
\newenvironment{proof}%
{\begin{trivlist}\item[\hspace*{\labelsep}{\it Proof.\/}]}%
{\hfill$\Box$\end{trivlist}}

\newtheorem{corollary}{Corollary}

\usepackage{latexsym,graphicx}
\begin{document}

\title{\bf Strip Packing vs. Bin Packing } %\thanks{Supported in part by }}
 
\author{Xin Han$^1$ \hspace{3mm} Kazuo Iwama$^1$\hspace{3mm}
        Deshi Ye$^2$
Guochuan Zhang$^{3}$
%%$^3$\thanks{Supported in part by the DFG Project AL
%%464/4-1, Eu-Project APPOL II and NSFC (10231060).}
\\ {\small $^1$ School of Informatics, Kyoto University, Kyoto
606-8501, Japan} \\ {\small \{hanxin,
iwama\}@kuis.kyoto-u.ac.jp}
\\ {\small $^2$ Department of Computer Science, The University of Hong Kong, Hong Kong}
\\ {\small  yedeshi@cs.hku.hk }
%%\\ {\small Georges-K\"ohler-Allee 79, 79110 Freiburg, Germany}
\\ {\small $^3$ Department of Mathematics, Zhejiang University, China}
\\ {\small zgc@zju.edu.cn}}
\date{}
\maketitle
%% \thispagestyle{headings}
%% \markright{Student paper}
\baselineskip 13.7pt

\begin{abstract}
%% Bin packing is one of the most extensively studied problems. A
%% more general problem is strip packing, in which one is asked to
%% pack a set of rectangles into a vertical strip of unit width so
%% that the total height of the strip used is minimum.
In this paper
we establish a general algorithmic framework between bin packing
and strip packing, with which we  achieve the same asymptotic
bounds by applying bin packing algorithms to strip packing. More
precisely we obtain the following results: (1) Any offline bin
packing algorithm can be applied to strip packing maintaining the same
asymptotic worst-case ratio. Thus using  FFD (MFFD) as a subroutine,
we get  a practical (simple and fast) algorithm 
for strip packing with an upper bound 11/9 (71/60).
A simple AFPTAS for strip
 packing immediately follows.
(2) A class of Harmonic-based
algorithms for bin packing can be applied to online strip packing
maintaining the same asymptotic competitive ratio. It implies
online strip packing admits an upper bound of 1.58889 
on the asymptotic competitive ratio,
which is very close to the lower bound 1.5401
and significantly improves
 the previously best bound of 1.6910 and affirmatively answers an
 open question posed \cite{cw97}.
\end{abstract}

\section{Introduction}
In strip packing a set of rectangles with widths and heights both
bounded by 1, is packed into a strip with width 1 and infinite
height. Rectangles must be packed such that no two rectangles
overlap with each other and the sides of the rectangles are parallel
to the strip sides. Rotations are not allowed. The objective is to
minimize the height of the strip to pack all the given rectangles.
If we know all rectangles before constructing a packing, then this
problem is {\em offline}. In contrast in {\em online} strip packing
rectangles are coming one by one and a placement decision for the
current rectangle must be done before the next rectangle appears.
Once a rectangle is packed it is never moved again.

It is well known that strip packing is a generalization of bin
packing. Namely if we restrict all input rectangles to be of the
same height, then strip packing is equivalent to bin packing. Thus
any negative results for bin packing still hold for strip packing.
More precisely, strip packing is NP-hard in the strong sense and the
lower bound 1.5401 \cite{{Vliet92}} is valid for online strip
packing.

\paragraph{\bf Previous results.} For the offline version Coffman
et al.~\cite{CGJT80} presented algorithms NFDH (Next Fit
Decreasing Height) and FFDH (First Fit Decreasing Height), and
showed that the respective asymptotic worst-case ratios are 2 and
1.7. Golan \cite{Golan81} and Baker et al. \cite{BBK81} improved
it to $4/3$ and $5/4$, respectively. Using linear programming and
random techniques, an asymptotic fully polynomial time
approximation schemes (AFPTAS) was given by Kenyon and R\'emila
\cite{KR00}. In the online version Baker and Schwarz \cite{BS83}
introduced an online strip packing algorithm called a shelf
algorithm. A shelf is a rectangular part of the strip with width
one and height at most one so that (i) every rectangle is either
completely inside or completely outside of the shelf and (ii)
every vertical line through the shelf intersects at most one
rectangle. Shelf packing is an elegant idea to exploit bin
packing algorithms. By employing bin packing algorithms {\em Next
Fit} and {\em First Fit} Baker and Schwarz \cite{BS83} obtained
the asymptotic competitive ratios of $2$ and $1.7$, respectively.
This idea was extended to the Harmonic shelf algorithm by Csirik
and Woeginger \cite{cw97}, obtaining an asymptotic competitive
ratio of $h_{\infty} \approx 1.6910$. Moreover it was shown that
$h_{\infty}$ is the best upper bound a shelf algorithm can
achieve, no matter what online bin packing algorithm is used.
Note that there were already several algorithms for online bin
packing that have asymptotic competitive ratios better than
$h_{\infty}$ in late 80s and early 90s
\cite{LL85,RBLL89,R91, Yao80}. Naturally an open question was posed
in \cite{cw97} for finding better online strip packing algorithms
that are not based on the shelf concept.

The core of shelf packing is reducing the two-dimensional problem to
the one-dimensional problem. Basically shelf algorithms consist of
two steps. The first one is {\em shelf design} which only takes the
heights of rectangles into account. One shelf can be regarded as a
bin with a specific height. The second step is {\em packing into a
shelf}, where rectangles with similar heights are packed into the
same shelves. This step is done by employing some bin packing
algorithms that pack the rectangles with a total width bounded 
 by one into a shelf. Clearly, to maintain the quality of bin
packing algorithms in shelf packing we must improve the first step.
Along this line we make the following contributions.

\paragraph {\bf Our contributions.} We propose a batch
packing strategy and establish a general algorithmic framework
between bin packing and strip packing. It is shown that any offline
bin packing algorithm can be used for offline strip packing
maintaining the asymptotic worst-case ratio. As an example, the well
known bin packing algorithm FFD can approximate strip packing with
an asymptotic worst-case ratio of 11/9. A simple AFPTAS can easily
be derived from \cite{kk82}.

We further prove that a class of online bin packing algorithm
based on Super Harmonic algorithm~\cite{s02} can be used in
online strip packing maintaining the same asymptotic competitive
ratio. This result implies that the known Harmonic based bin
packing algorithms \cite{LL85,RBLL89,R91,s02} can be converted
into online strip packing algorithms without changing their
asymptotic competitive ratios (better than $h_\infty$), and thus
affirmatively answers the open question in \cite{cw97}. Note that
the current champion algorithm for online bin packing is
Harmonic++ by Seiden \cite{s02}, which has an asymptotic
competitive ratio of 1.58889. Hence strip packing admits an online
algorithm with the same upper bound of 1.58889.

\paragraph{\bf Main ideas.} Recall that strip packing
becomes bin packing if all rectangles have the same height. It
motivates us to construct new rectangles with the same height by
bundling a subset of given items. More precisely, in the offline
case, we pack in batch the rectangles with similar width into
rectangular bins of pre-specified height of $c$,
 where $c   >1$ is a
sufficiently large constant. Then we obtain a set of new
rectangles (rectangular bins) of the same height. The next step is
to use  bin packing algorithms on the new set. In the
on-line case the strategy is slightly different. We divide the
rectangles into two groups according to their widths, to which we
apply the above batching strategy and the standard shelf
algorithms respectively.

%% \paragraph {\bf Remarks on guillotine cuts.}
%% In stock-cutting applications, machines can only perform
%% edge-to-edge cuts parallel to the strip's edge, that are called
%% guillotine cuts. And each stage consists of either horizontal or
%% vertical guillotine cuts (but not both). If any rectangle can be
%% obtained from the strip by at most $k$ guillotine cuts including a
%% possible cut to separate the rectangle itself from a waste area,
%% then we say the cutting (or packing) is $k$-stage of guillotine
%% cuts. Figure~\ref{fig:slips} shows an illustration: (a) is 3-stage
%% of guillotine cuts, while (b) is 4-stage of guillotine cuts.

%% The AFPTAS in \cite{KR00} gives 5-stage of guillotine cuts. Seiden
%% and Woeginger \cite{SW05} recently revisited the problem and
%% presented an AFPTAS with 4-stage guillotine cuts. Moreover they
%% proved that with only 3-stage guillotine cuts the asymptotic
%% worst-case ratio cannot be better than $h_\infty$. In this paper all
%% algorithms derived are 4-stage of guillotine cuts.

\paragraph {\bf Asymptotic worst-case (competitive) ratio.} To evaluate an approximation
(online) algorithms for strip packing and bin packing we use the
standard measure defined as follows.

Given an input list {\em L} and an approximation (online) algorithm
$A$, we denote by $OPT(L)$ and $A(L)$, respectively, the height of
the strip used by an optimal (offline) algorithm and the height used
by (online) algorithm $A$ for packing list $L$.

The {\em asymptotic worst-case (competitive) ratio} $R_A^{\infty}$
of algorithm $A$ is defined by
\[
     R_A^{\infty} =\lim_{n \to \infty} \sup_{L}\{ A(L)/OPT(L)| OPT(L) = n\}.
\]

\section{The offline problem}
Given a rectangle $R$, throughout the paper, we use $w(R)$ and
$h(R)$ to denote its width and height, respectively.

\vskip 2mm\noindent{\bf Fractional strip packing.} A fractional
strip packing of $L$ is a packing of any list $L'$ obtained from $L$
by subdividing some of its rectangles by horizontal cuts: each
rectangle $(w,h)$ is replaced by a sequence
$(w,h_1),(w,h_2),...,(w,h_k)$ of rectangles such that $h =
\sum_{i=1}^{k}h_i$.
%%Note that our definition is slightly different from the one in \cite{KR00}.

\vskip 2mm\noindent{\bf Homogenous lists.} Let $L$ and $L'$ be two
lists where any rectangle of $L$ and $L'$ takes a width from $q$
distinct numbers $w_1>w_2>\cdots>w_q$. List $L$ is {\em
$r$-homogenous} to $L'$ where $r\ge 1$ if
$$\sum\limits_{w(R^{'})=w_i,
R'\in L'} h(R^{'})\le \sum\limits_{w(R)=w_i,R\in L}h(R) \le r\cdot
\sum\limits_{w(R^{'})=w_i, R'\in L'} h(R^{'}).$$

The following lemma is an implicit byproduct of the APTAS for strip
packing given by Kenyon and R\'emila \cite{KR00}.
%%(The proof will be appeared in a complete version.)
\begin{lemma} \label{lemma:fractional}
For each strip packing instance $I$ and $\epsilon > 0$, we have
$OPT(I) \le (1+\epsilon)OPT_{FSP}(I) + O(\epsilon^{-2})$, where
$OPT_{FSP}(I)$ is the optimal value of fractional strip packing for
instance I.
\end{lemma}

%% \begin{proof}
%%  Sketch.
%%  This lemma can be obverstion as below:

%%  i) divide the items into two groups, say {\em wide} and {\em narrow},
%%     then round up  widths of all {\em large} items into  distinct values
%%     such that this rounding does not change the optimal solution significantly.
 
%%  ii) use LP relaxation  to get a feasible solution for {\em wide} items.

%%  iii) pack {\em narrow} items by NFDH packing.

%%  By the above approach, this lemma follows.
%% \end{proof}
The next lemma shows a useful property of {\em homogenous} lists.
\begin{lemma} \label{lemma:equal}
Given two lists $L$ and $ L'$, if $L$ is $r$-homogenous to $L'$,
we have $OPT_{FSP}(L')\le OPT_{FSP}(L) \le r\cdot OPT_{FSP}(L')$.
%where
%$OPT_{FSP}(\cdot$) denotes the optimal fractional strip packing's value.
\end{lemma}
\begin{proof}
If $r=1$, it is easy to see that any fractional strip packing of $L$
is a fractional packing of $L'$ and vice versa. The conclusion thus
follows immediately.

Now we consider the case that $r>1$. By adding some rectangles to
$L'$ we can get a new list $L'_1$ which is $1$-homogenous to $L$. We
have 
\[
OPT_{FSP}(L')\le OPT_{FSP}(L'_1)= OPT_{FSP}(L).
\]
 On the other hand we obtain another list $L'_2$ by prolonging in height all
rectangles of $L'$, i.e., if $(w,h)\in L'$, then $(w,rh)\in L'_2$.
Clearly 
\[
OPT_{FSP}(L'_2)\le r\cdot OPT_{FSP}(L').
\]
 Moreover, $OPT_{FSP}(L)\le OPT_{FSP}(L'_2)$. The lemma holds.
\end{proof}

\begin{theorem}\label{th:near}
Given two lists $L$ and $ L'$, if $L$ is $r$-homogenous to $L'$,
then for any $\epsilon>0$
\[
 OPT(L) \le r(1+\epsilon)OPT(L') + O(\epsilon^{-2}).
\]
\end{theorem}
\begin{proof} By Lemma \ref{lemma:fractional},
$$
     \begin{array}{c}
       OPT(L) \le (1+\epsilon)OPT_{FSP}(L) + O(\epsilon^{-2}). %\\
       %OPT(L') \le (1+\epsilon)OPT_{FSP}(L') + O(\epsilon^{-2}).
     \end{array}
$$
By Lemma \ref{lemma:equal},
  \[
     OPT_{FSP}(L)\le r\cdot OPT_{FSP}(L').
  \]
Moreover %$OPT_{FSP}(L) \le OPT(L)$ and
$OPT_{FSP}(L') \le OPT(L')$.
Hence we have this theorem.
\end{proof}
In the following we are ready to present our approach for offline
strip packing. Given an input list $L=\{ R_1, \dots\ ,R_n\}$ such
that $w_1 \ge w_2 \ge \cdots \ge w_n$, where $R_i=(w_i, h_i)$, and a
constant $c>1$, we construct an offline algorithm $B\&P_A$ using
some bin packing algorithm $A$ as a subroutine. Basically the
strategy consists of two stages.

\begin{figure}[htbp]
\begin{center}
\includegraphics[scale=0.5]{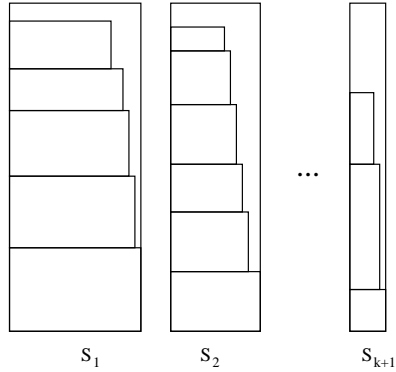}
\caption{ Packing rectangles into slips} \label{fig:FFDW}
\end{center}
\end{figure}

\vskip 2mm\noindent {\bf Stage 1} - $Batching$. Pack $R_1,\dots,R_i$
by NF algorithm in the vertical direction into a slip $S_1=(w_1,c)$,
where $\sum_{j=1}^{i}h_j \le c < \sum_{j=1}^{i+1}h_j $ and pack
$R_{i+1},\dots,R_{k}$ into a slip $S_2=(w_{i+1},c)$, and so on,
until all items are packed, shown as Figure~\ref{fig:FFDW}. (Note
that except for the last slip, all slips have the packed heights  at
least $(c-1)$.)

\vskip 2mm\noindent {\bf Stage 2} - $Packing$. Except for the last
slip, pack all slips into the strip by algorithm $A$, since all
slips have the same heights $c$. Then append the last slip on the
top of the strip.

\vskip 2mm We present the main result for the offline case. In
terms of the asymptotic worst case ratio, strip packing is
essentially the same as bin packing.
\begin{theorem}\label{theo:general}
The asymptotic worst-case ratio $R_{B\&P_A}^{\infty}=R_{A}^{\infty}$
for any bin packing algorithm $A$.
\end{theorem}
\begin{proof} Assume that $R_{A}^{\infty} = \alpha$.
After the first stage of algorithm $B\&P_A$, we get a series of
slips $S_1,\ldots, S_k, S_{k+1}$, shown as Figure~\ref{fig:FFDW}. We
then round up every item $(w_j, h_j)$ in slip $S_i$ to
$(w(S_i),h_j)$ and obtain a new list $\bar{L}$, where $w(S_i)$ is
the width of slip $S_i$. On the other hand, we obtain another list
{\em \underbar{L}} by rounding down every item $(w_j, h_j)$ in slip
$S_i$ to $(w(S_{i+1}),h_j)$ (here we set $w(S_{k+2}) =0$). We have
\begin{equation}
   OPT(\underbar{L}) \le OPT( L) \le OPT(\bar{L})
  \label{equation:rounding1}
\end{equation}
Denote two sets $L_1=\{S_1,\dots,S_k\}$ and $L_2=\{S_2,\dots,S_k\}$.
Then
   \begin{equation}
   OPT(L_2) \le OPT( L_1) \le OPT( L_2) + c.
  \label{equation:rounding2}
 \end{equation}
We can treat $S_i$ as a one-dimensional item ignoring its height
since $h(S_i)=c$ for $i=1,2,\ldots,k $. Let $I(L_1)$ be the
corresponding item set for bin packing induced from the list $L_1$,
i.e, $I(L_1)=\{w(S_1),w(S_2),\ldots, w(S_k)\}$. And $OPT(I(L_1))$ is
the minimum number of bins used to pack $I(L_1)$. It follows that
$OPT( L_1)=c\cdot OPT(I(L_1))$.

Note that $L_2$ is $c/(c-1)$-homogenous to {\em \underbar{L}}, by
Theorem \ref{th:near}, we have
\begin{equation} \label{equation:near}
OPT(L_2) \le \frac{c}{c-1}(1+\epsilon)OPT(\underbar{L}) +
O(c+\epsilon^{-2}).
\end{equation}
Now we turn to algorithm $B\&P_A$. After Stage 1 the list $L$
becomes $L_1\cup \{S_{k+1}\}$. At Stage 2 we deal with a bin packing
problem: pack $k+1$ items with size of $w(S_i)$ into the minimum
number of bins. The bin packing algorithm $A$ is applied to $I(L_1)$
while $S_{k+1}$ occupies a bin itself. Thus $B\&P_A(L) \le c\cdot
A(I(L_1)) + c$. Since $R_{A}^{\infty}=\alpha$, we have $A(I(L_1))
\le \alpha OPT(I(L_1))+ O(1)$. Then $$B\&P_A(L) \le c\cdot A(I(L_1))
+ c \le \alpha\cdot c\cdot OPT(I(L_1)) + O(c)=\alpha\cdot OPT(L_1) +
O(c).$$ Combining with
(\ref{equation:rounding2}),(\ref{equation:near}),
(\ref{equation:rounding1}), we have
\begin{eqnarray}
      B\&P_A(L) & \le& \alpha OPT(L_2) + O(c)   \\
              & \le&\frac{\alpha c}{(c-1)} (1+\epsilon) OPT(\underbar{L}) + O(\epsilon^{-2} +c)  \\
              & \le& \frac{\alpha c}{(c-1)} (1+\epsilon) OPT(L) + O(\epsilon^{-2}+c).
\end{eqnarray}
As $c$ goes to infinite, this theorem follows.
\end{proof}
By Theorem \ref{theo:general}, any offline bin packing algorithm
can be transformed into an offline strip packing algorithm without
changing the asymptotic worst case ratio.
 If the well known algorithm FFD (\cite{B85} \cite{J73}\cite{Y91}) is used 
 in our approach, 
 then we get a simple and fast algorithm $B\&P_{FFD}$ for strip
packing and have the following result from Theorem
\ref{theo:general}.
\begin{corollary}
Given constants  $\epsilon >0$ and $c >1$, for any strip packing
instance $L$, $B\&P_{FFD}(L)\le \frac{11c}{9(c-1)} (1+\epsilon)
OPT(L) + O(\epsilon^{-2} +c)$, where $c \le \epsilon OPT(L)$.
\end{corollary}

\section{The online problem}
In this section we consider online strip packing. In the online case
we are not able to sort the rectangles in advance because of no
information on future items. Due to this point we cannot reach a
complete matching between bin packing algorithms and strip packing
algorithms generated from the former. However we can deal with a
class $H$ of Super Harmonic algorithms \cite{s02}(to be given in the appendix),
which includes all known online bin packing algorithms based on Harmonic.
Such an algorithm can be used in online strip packing without
changing its asymptotic worst-case ratio.

A general algorithm of Super Harmonic  algorithms has the
following characteristics.
\begin{itemize}

\item Items are classified into $k+1$ groups by their sizes,
      where $k$ is a constant integer.

\item Those items in the same group are packed by the same manner.

\end{itemize}

Let $A$ be any algorithm of Super Harmonic  algorithm. Our
approach $G\&P_A$ is presented below.

\vskip 2mm \noindent{\bf Grouping:} A rectangle is {\em wide} if its
width is at least $\epsilon$; otherwise it is {\em narrow}, where
$\epsilon >0$ is a given small number. We further classify {\em wide}
rectangles into $k$ classes, where $k$ is a constant, as Algorithm
$A$ does. Let $1=t_1>t_2>\cdots>t_{k}>t_{k+1}= \epsilon$. Denote $I_j$
to be the interval $(t_{j+1}, t_j]$ for $j= 1,...,k$. A rectangle is
of type-$i$ if its width $w \in I_i$.

\vskip 2mm\noindent{\bf Packing narrow rectangles:} Apply the
standard shelf algorithm $NF_r$ \cite{BS83} to {\em narrow}
rectangles $R=(w,h)$, where $0<r<1$ is a parameter. Round $h$ to
$r^s$ if $r^{s+1} < h \le r^s$. If $R$ cannot be packed into the
current open shelf with height of $r^s$, then close the current one
and open a new one with height $r^s$ and pack $R$ into it, otherwise
just pack $R$ into the current one by NF.

\begin{figure}[htbp]
  \begin{center}
  \includegraphics[scale=0.7]{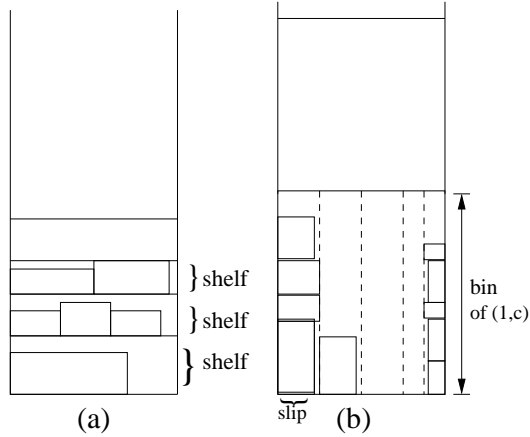}
  \caption{ Shelf packing vs our packing}
  \label{fig:slips}
  \end{center}
\end{figure}

\vskip 2mm\noindent{\bf Packing wide rectangles:} We pack {\em wide}
rectangles into bins of $(1,c)$, where $c = o(OPT(L))>1$ is a constant.
Similarly as the offline case we batch the items of the same type
and pack them into a slip. Here we specify the width of the slip by
values $t_i$ for $i<k+1$ and name a slip $(t_i,c)$ of type-$i$.
Suppose that the incoming rectangle $R$ is of type $i$ ($w \in
(t_{i+1},t_i]$). If there is a slip of type-$i$ with a packed height
less than $c-1$, then pack $R$ into it by algorithm NF in the
vertical direction. Otherwise create a new empty slip of type-$i$
with size $(t_i,c)$ and place $R$ into the new slip by NF algorithm
in the vertical direction. As soon as a slip is created,
view it as one dimensional item and  pack it by
algorithm $A$ into a bin of $(1,c)$. Figure~\ref{fig:slips}(b) shows
an illustration.

The weighting function technique introduced by Ullman
\cite{ullman71} has been widely used in performance analysis of bin
packing algorithms \cite{cw97}\cite{LL85}\cite{s02}. Roughly
speaking, the weight of an item indicates the maximum portion of a
bin that the item  occupies.
Then, Seiden generalized the idea of weighting function
and proposed  a weighting system which can be used to analyze 
Harmonic, Refined Harmonic, Modified Harmonic, 
Modified Harmonic 2, Harmonic+1 and Harmonic++.
The following analysis of $G\&P_A$ is based on
the weighting system proposed by Seiden \cite{s02}.
%%first we introduce it and give some basic properties based it.
  
{\bf Weighting Systems}:
 Let $\mathbb{R}$ and $\mathbb{N}$ be the sets of real numbers and
 nonnegative integers, respectively.
 A {\em weighting system} for algorithm $A$ is a tuple
 $(\mathbb{R}^{m}, {\bf w}_{A},\xi_{A})$.
 $\mathbb{R}^{m}$ is a vector space over the real numbers with
 dimension $m$.
 The function ${\bf w}_{A}:(0,1] \mapsto \mathbb{R}^{m}$ is called
 the {\em weighting function}.
 The function $\xi_{A}: \mathbb{R}^{m}\mapsto  \mathbb{R}$ 
 is called the {\em consolidation function}.
 Seiden defined a $2K+1$ dimensional weighting system for Super Harmonic,
 where $K$ is a parameter of Super Harmonic algorithm.
 Real numbers $\alpha_i, \beta_i, \gamma_i, \epsilon$ and functions
 $\phi(i), \varphi(i)$ are defined in Super Harmonic algorithm.
 The unit basis vectors of the weighting system are denoted by
 \[
  {\bf b}_0, {\bf b}_1,....,{\bf b}_{K},{\bf r}_1,....,{\bf r}_K.
\] 
 The weighting function is 
 \[
    {\bf w}_{A}(x) = \left\{
                     \begin{array}{ll}
                      (1-\alpha_i)\frac{{\bf b}_{\phi(i)}}{\beta_i} + 
                      \alpha_i \frac{{\bf r}_{\varphi(i)}}{\gamma_i} &
                      \textrm{ if   $x \in I_i$ with $i \le k$,} \\
                      x\frac{{\bf b}_0}{1-\epsilon}   &  
                      \textrm{ if $x \in I_{k+1}$.}
                     \end{array} 
                     \right.
 \]
 The consolidation function is 
  \[
   \xi_{A}({\bf x}) = {\bf x} \cdot {\bf b}_0 + \max_{ 1 \le j \le K+1} 
     \min \Big\{  \sum_{i=j}^{K}{\bf x} \cdot {\bf r}_i + 
                   \sum_{i=1}^{K}{\bf x} \cdot {\bf b}_i, 
                    \sum_{i=1}^{K}{\bf x} \cdot {\bf r}_i +
                      \sum_{i=1}^{j-1}{\bf x} \cdot {\bf b}_i  \Big\}.
\]
\begin{lemma}\cite{s02} 
 For all sequences of bin packing $\delta =(p_1,...,p_n)$,
 \[
   cost_A(\delta) \le \xi_A \Big ( \sum_{i=1}^n {\bf w}_A(p_i)\Big ) + O(1).
\]
 \label{lemma:cost}
\end{lemma}
This means that the cost of algorithm $A$
is bounded by the total weight of the items.

We can obtain a similar result with  Lemma \ref{lemma:cost}
by defining our weighting function as follows,
\[
   {\bf w}_{A}(P) = y \cdot {\bf w}_{A}(x), 
\]
where  P is a rectangle of size $(x,y)$.

%% We use a similar weighting system for $G\&P_A$ as the one
%% for Super Harmonic, but the {\em weighting function} is 
%%  a little different.
%% Given a rectangle $P=(x,y)$,
%% our weighting function is defined as
%% \[
%%    {\bf w}_{A}(P) = y \cdot {\bf w}_{A}(x). 
%% \]
%%And we  have the similar result with the one in  lemma \ref{lemma:cost}.
\begin{lemma}
 For any sequence of rectangles $L=(P_1,...,P_n)$,
 the cost by $G\&P_A$ is 
\[
   cost_A(L) \le \max\{\frac{c}{c-1},\frac1r\}
 \xi_A \Big ( \sum_{i=1}^n {\bf w}_A(P_i)\Big ) + O(1).  
\]
 \label{lemma:ourcost}
\end{lemma}
Since the proof is similar with the one in \cite{s02},
we give it in the appendix.

For bin packing, a {\em pattern} is a tuple $q = \langle q_1,...,q_k \rangle$ 
over $\mathbb{N}$ such that
\[
        \sum_{i=1}^{k} q_i t_{i+1} < 1,
\]
where $q_i$ is the number of items of type $i$ contained in the bin.
Intuitively, a pattern describes the contents of a bin.
The weight of pattern $q$ is
\[
    {\bf w}_A(q) = {\bf w}_A\Big(1- \sum_{i=1}^{k} q_i t_{i+1} \Big) 
                 +  \sum_{i=1}^{k} q_i {\bf w}_A(t_i).
\]
Define $\mathcal{Q}$ to be the set of all patterns $q$. 
Note that $\mathcal{Q}$ is necessarily finite.

A {\em distribution} is a function 
$\chi : \mathcal{Q} \mapsto \mathbb{N}_{\ge 0}$ such that
\[
  \sum_{q \in \mathcal{Q} } \chi(q) = 1.
\]
Given an instance of bin packing $\delta$,
Super Harmonic uses cost($\delta$)$\chi(q)$ bins
containing items as described by the pattern $q$.
\begin{lemma}\cite{s02}
 For any distribution $\chi$, if we set $A$ as Harmonic++ then
 \[
   \xi_A \Big( \sum_{q \in \mathcal{Q}} \chi(q) {\bf w}_A(q)\Big)
   \le 1.58889.
\]
\label{lemma:1.58889}
\end{lemma}
\begin{theorem}
 If we set algorithm $A$ to Harmonic++,
 then the asymptotic competitive ratio of algorithm $G\&P_A$ is 
 $1.58889$, where $c$ is a constant..
 \label{theorem:1.58889}
\end{theorem}
\begin{proof}
Given an optimal packing for $L$,
we cut the optimal packing into layers such that 
all rectangles in each layer have the same height, 
shown as in Fig.~\ref{fig:cutting}.
(Here the rectangle may be a part of the original one.)
\begin{figure}[tbp]
  \begin{center}
  \includegraphics[scale=0.7]{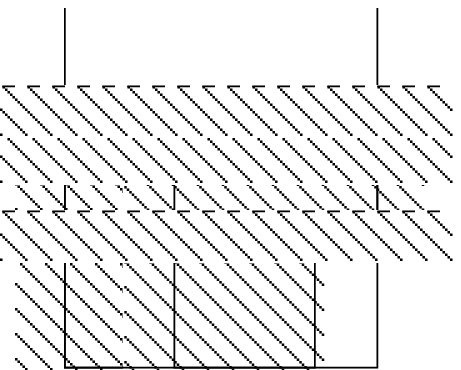}
  \caption{ Cutting an optimal packing into layers}
  \label{fig:cutting}
  \end{center}
\end{figure}

Now, we show this cutting does not change the total weight.
Given a rectangle $R=(x,y)$,
if we cut it into $P_1,...,P_m$ such that $P_i=(x,y_i)$ and
$y=\sum_i y_i $,
then 
\[
{\bf w}_A(R) = y{\bf w}_A(x) = \sum_{i} y_i {\bf w}_A(x) = \sum_{i} {\bf w}_A(P_i).
\]
Let $L'$ be the list induced from $L$ by the above cutting.
Then
\begin{equation}
  \xi_A\Big(\sum_{R \in L}{\bf w}_A(R)\Big) 
= \xi_A\Big(\sum_{R \in L'}{\bf w}_A(R)\Big).
 \label{eqn:cutting}
\end{equation} 
It is not difficult to see each layer corresponds to a {\em pattern} of
bin packing. 
Let $h_q$ is the total height of the pattern $q$.
So, 
\[
 OPT(L)=\sum_{q \in \mathcal{Q}} h_q = \sum_{q \in \mathcal{Q}} OPT(L) \chi(q),
\]
where $\mathcal{Q}$ is the set of all pattern and $\chi(q)$ is one
distribution of  $\mathcal{Q}$.
Then
 \[
 \xi_A\Big(\sum_{R \in L'}{\bf w}_A(R)\Big) \le 
 \xi_A\Big(  \sum_{q \in \mathcal{Q} }h_q {\bf w}_A(q) \Big)
  = OPT(L)\xi_A\Big(  \sum_{q \in \mathcal{Q}} \chi(q) {\bf w}_A(q)\Big). 
\]
If we set algorithm $A$ to Harmonic++,
then by lemma \ref{lemma:1.58889},
$\xi_A\Big(\sum_{q \in \mathcal{Q}} \chi(q) {\bf w}_A(q)\Big) \le 1.58889$,
then by (\ref{eqn:cutting}),
 $ \xi_A(\sum_{R \in L}{\bf w}_A(R)) \le 1.58889OPT(L)$.
By lemma \ref{lemma:ourcost} and when $r$ goes to 1 and $c$ goes to $\infty$,
the asymptotic competitive ratio of algorithm $G\&P_A$ is $1.58889$.
\end{proof}

\section{Concluding Remarks}
Although strip packing is a generalization of the one dimensional
bin packing problem, we show from the point of algorithmic view
that it is essentially the same as bin packing. In terms of
asymptotic performance we give a universal method to apply the
algorithmic results for bin packing to strip packing maintaining
the solution quality. However our approach cannot be applied to
strip packing in terms of absolute performance. Note that
algorithm FFD has an absolute worst-case ratio of $3/2$ which is
the best possible unless $P=NP$. It is challenging to prove or
disprove the existence of a 3/2-approximation algorithm for
offline strip packing.

\normalsize
\newpage
\begin{center}
{\bf Appendix}
\end{center}
%% \subsection{Harmonic Shelf Algorithm}
%% So far,  for online strip packing, the best asymptotic competitive
%% ratio 1.6910, comes from Harmonic Shelf (HS) algorithm \cite{cw97}.
%% In HS, all rectangles are first grouped by their heights. Then the
%% packing is constructed as a sequence of shelves, and rectangles
%% belonging to the same group are packed into  shelves with the same
%% heights. In each shelf, Harmonic algorithm \cite{LL85} is used. To
%% be more precise, whenever  a new rectangle $(w_i, h_i)$ arrives, the
%% shelf algorithm determines the integer $s$ such that $r^{s+1} < h_i
%% \le r^{s} $ and puts it into class $C_s$, where $r <1$ is a positive
%% real. All rectangles in $C_s$ are packed into shelves of height
%% $r^{s}$ by Harmonic algorithm \cite{LL85}.

%% Note that in paper \cite{cw97}, they claimed that any shelf
%% algorithm for online strip packing has a lower bound  $h_{\infty}
%% \approx 1.6910$, if the {\it shelf algorithm}  satisfies the two
%% following conditions: i) every rectangle is either completely inside
%% or completely outside of a shelf, ii) every vertical line through
%% the shelf intersects at most one rectangle, shown as
%% Figure~\ref{fig:slips} (a). Actually, the lower bound 1.6910 in
%% \cite{cw97} is  the lower bound for any 2-stages of guillotine cuts.
\noindent{\bf Super Harmonic Algorithm}

In Super Harmonic \cite{s02} algorithm, items are classified into $k+1$
classes, where $k = 70$. Let $t_1 = 1 > t_2 >...>$ $t_{k} > t_{k+1}
= \epsilon > t_{k+2}=0 $ be real numbers. The  interval $I_j$ is
defined to be $(t_{j+1}, t_j]$ for  $j= 1,...,k$. And an item with
size $x$ has type-i if $x \in I_i$.
 %%Note that these intervals are disjoint and that they  cover $(0,1]$.

{\bf Parameters in Harmonic algorithm:} Each  type-i item  is
assigned a color, {\em red} or {\em blue}, $i \le k$.

The algorithm uses two sets of counters, $e_1,...,e_k$ and
$s_i,...,s_k$, all of which are initially zero. The total number of
type-i items is denoted by $s_i$, while the number of type-i {\em
red} items is denoted by $e_i$. For $1 \le i \le k$, during packing
process, for type-i items,  the balance between $s_i$ and $e_i$ is
kept, i.e., $e_i=\lfloor \alpha_i s_i \rfloor$, where
$\alpha_1,...,\alpha_k \in$ [0,1] are constants.

$\delta_i = 1 -t_i\beta_i$ is  the left space when a bin is filled
with $\beta_i$ type-i items. If possible, the left space is used for
{\em red} items.
 %% However, we sometimes use the width less than $\delta_i$ in order to simplify
%%  the algorithm and its analysis.
$D=\{\Delta_1,...,\Delta_K\}$ is the set of spaces into which red
items can be packed, and $0 = \Delta_0 < \Delta_1 < \cdots <
\Delta_{K} < 1/2$; where $K \le k$. $\Delta_{\phi(i)}$ is the  space
used to hold {\em red} items in a bin which holds $ \beta_i$ {\em
blue} items of type-i, where function $\phi$ is defined as
\{1,...,k\} $\mapsto$ \{0,...,K\}. And $\phi$ satisfies
$\Delta_{\phi(i)} \le \delta_{i}$. $\phi(i) = 0$ indicates that no
red items are accepted. Define $\gamma_i=0$ if $t_i > \Delta_{K}$,
otherwise $\gamma_{i}= max\{1,\lfloor \Delta_1/t_i \rfloor\}$.
In the case that $\Delta_K \ge t_i > \Delta_1$,
we set $\gamma_i = 1$.
Again, this seems to be the best choice from a worst case perspective.
Define
\[
  \varphi(i) = \min\{j |  t_i \le \Delta_j, 1 \le j \le K \}.
\]
Intuitively, $\varphi(i)$ is the index of the smallest space in $D$ 
into which a red item of type i can be placed.

{\bf Naming bins:}  Bins are named as follows:
 \[
    \{i|\phi_i = 0, 1 \le i \le k,\}
 \]
 \[
   \{(i,?)|\phi_i \ne 0, 1 \le i \le k,\}
  \]
 \[
   \{(?,j)|\alpha_j \ne 0, 1 \le j \le k,\}
  \]
\[
    \{(i,j)|\phi_i \ne 0, \alpha_j \ne 0, \gamma_j t_j \le \Delta_{\phi(i)},
1 \le i,j \le k\}.
 \]
 %% \{i\}, \{i,?\}, \{?,j\}, \{i,j\} respectively,
%%   where $1 \le i,j \le k$.
Group $(i)$ contains bins that hold only {\em blue} items of type-i.
Group $(i,j)$ contains bins that contain {\em blue} items of type-i
and {\em red} items of type-j.
 %% and a filled bin $(i,j)$ contains $\beta_{i}$ items .
{\em Blue} group $(i,?)$ and {\em red} group $(?,j)$ are
indeterministic bins, in which they {\em currently} contain only
{\em blue} items of type-i or {\em red} items of type-j
respectively. During packing, {\em red} items or {\em blue} items
will be packed if necessary, i.e., indeterministic bins will be
changed into $(i,j)$.

 {\bf Super Harmonic }
   \begin{enumerate}
%%   \item  Initialize $e_i \gets 0$ and $s_i \gets 0$ for $1 \le i \le k$.
   \item  For each item $p$:  $i \gets$ type of $p$,
        \begin{enumerate}
          \item if $i = k+1$ then using NF algorithm,
           \item else   $s_i \gets s_i +1$;
             if $e_i < \lfloor \alpha_i s_i \rfloor$
                  then $e_i \gets e_i +1$; \{ color p  red \}
                \begin{enumerate}
%%                  \item $e_i \gets e_i +1$; \{ color p  red \}
                  \item If there is a bin in group $(?,i)$ with fewer
                   than $\gamma_i$ type-i items,
                   then place p in it.  \\
                   Else if, for any $j$, there is a bin in group $(j,i)$
                   with fewer than $\gamma_i$  type-i items then
                   place p in it.
                   \item  Else if there is some bin in group $(j,?)$ such
                         that $\Delta_{\phi(j)} \ge \gamma_i t_i $, then
                          pack $p$ in it and
                        change  the bin into $(j,i)$.
                  \item Otherwise, open a bin $(?,i)$, pack $p$ in it.
                \end{enumerate}
          \item else \{color p blue\}:
             \begin{enumerate}
             \item if $\phi_i =0$ then
 %                  \begin{enumerate}
                       if there is a bin  in group $i$ with
                            fewer than $\beta_i$ items then pack $p$ in it,
                            else  open a new group $i$ bin,
                           then pack $p$ in it.
 %                  \end{enumerate}
              \item Else:
                   \begin{enumerate}
                    \item if, for any $j$, there is a bin in
                         group $(i,j)$  or $(i,?)$
                         with fewer than $\beta_i$ type-i items,
                         then pack $p$ in it.
%%                          Else if there is an open bin in group $(i,?)$ with fewer
%%                          than $\beta_i$ type-i items,
%%                           then pack $p$ in it.

                    \item  Else if there is a bin in group $(?,j)$ such
                           that $\Delta_{\phi(i)} \ge \gamma_j t_j$ then
                              pack $p$ in it,
                            and change the group of this bin into $(i,j)$.
                    \item Otherwise, open a new bin $(i,?)$ and
                           pack $p$ in it.
                     \end{enumerate}
             \end{enumerate}
      %%       \end{enumerate}
        \end{enumerate}
   \end{enumerate}

\begin{lemma}
 If the total area of {\em narrow} rectangles is $S$ then
 the cost for {\em narrow} rectangles
 by $G\&P_A$ is at most $\frac{S}{r(1-\epsilon)} +O(1)$. 
 \label{lemma:shelfpacking}
\end{lemma}

\begin{proof}
 Note that every narrow rectangle has  its width at most $\epsilon$.
 Given a close shelf with height $h$,
 the total area of rectangles in it is larger than $r \cdot h(1-\epsilon)$.
 If the total cost of all close shelves is $H_1$,
 then $S > r\cdot H_1(1-\epsilon)$.
 On the other hand,
 at any time in the strip packing maintained by algorithm $G\&
 P_A$, the total cost of all open shelves (in each of which the
 total width of rectangles packed is less than $1-\epsilon$)
 is less than $\sum_{i=0}^\infty r^i= 1/(1-r)$ ($0<r<1$).
 
 So the total  cost for narrow items
 is at most $\frac{S}{r(1-\epsilon)} +O(1)$. 
\end{proof}

\begin{lemma}
 $G\&P_A$ algorithm maintains the following invariants.

 i) at most one bin has fewer than $\beta_i$ slips in any group
 $(i,?)$ or $(i)$.
 
 ii) at most one bin has fewer than $\gamma_i$ slips in any group
 $(?,i)$.

 iii) at most three bins have fewer than  $\beta_i$ slips
      or   fewer than $\gamma_i$ slips in any group $(i,j)$.

iiii) at most $k$ bins have a slip with the total packed height
      less than $c-1$.
\label{lemma:openbins}
\end{lemma}
\begin{proof}
  Since i), ii), iii) are direct from Lemma 2.2 \cite{s02},
  we just prove the claim in iiii).
 Totally, there are $k$ kinds of slips and we maintains that
 any time, for each kind of a slip, there is at most one slip
 with the total packed height less $c-1$.
 So, iiii) holds.
\end{proof}

In Super Harmonic algorithm,
if we define the {\em class} of a red item of type $i$ to be $\varphi(i)$
and the class of a blue item of type $i$ to be $\phi(i)$.
Let $B_i$ and $R_i$ be the number of bins containing blue items
of class $i$ and red items of class $i$, respectively.
\begin{lemma}\cite{s02} 
 In Super Harmonic algorithm, the total number
 of bins for red and blue items is at most
 \[
 B_0   + \max_{ 1 \le j \le K+1} 
     \min \Big\{  \sum_{i=j}^{K} R_i + 
                   \sum_{i=1}^{K} B_i, 
                    \sum_{i=1}^{K} R_i +
                      \sum_{i=1}^{j-1} B_i  \Big\} + O(1).
\]
 \label{lemma:redblue}
\end{lemma}

\paragraph{ Proof of Lemma \ref{lemma:ourcost}}
 \begin{proof}
Let $B_i$ and $R_i$ be the number of bins containing blue slips
of class $i$ and red slips of class $i$, respectively.
Let $D$ be the total area of narrow rectangles.
By lemmas \ref{lemma:shelfpacking}, \ref{lemma:openbins}, \ref{lemma:redblue},
the total cost $cost_A(L)$ is at most
\begin{eqnarray*}
 & &\frac{D}{r(1-\epsilon)} + c \cdot \Big( B_0 +   \max_{ 1 \le j \le K+1} 
      \min \Big\{  \sum_{i=j}^{K} R_i + 
                   \sum_{i=1}^{K} B_i, 
                    \sum_{i=1}^{K} R_i +
                      \sum_{i=1}^{j-1} B_i  \Big\} \Big)   + O(1) 
%%  &\le& \max\{\frac1r, \frac{c}{c-1} \} 
%%   \Big\{ \frac{D}{1-\epsilon}  + \max_{ 1 \le j \le K+1} 
%%      \min \Big\{  \sum_{i=j}^{K} R_i + 
%%                    \sum_{i=1}^{K} B_i, 
%%                     \sum_{i=1}^{K} R_i +
%%                       \sum_{i=1}^{j-1} B_i  \Big\} \Big\} + O(1) \\
%%  &=& \max\{\frac1r, \frac{c}{c-1} \} 
%%   \Big\{ B_0  + \max_{ 1 \le j \le K+1} 
%%      \min \Big\{  \sum_{i=j}^{K} R_i + 
%%                    \sum_{i=1}^{K} B_i, 
%%                     \sum_{i=1}^{K} R_i +
%%                       \sum_{i=1}^{j-1} B_i  \Big\} \Big\} + O(1) \\
%% &=& \max\{\frac1r, \frac{c}{c-1} \} cost_A(\delta) \\
%% &\le &  \max\{\frac{c}{c-1},\frac1r\}
%%  \xi_A \Big ( \sum_{i=1}^n {\bf w}_A(R_i)\Big ) + O(1),
\end{eqnarray*}

To complete the proof, we show that this is at 
$\max\{\frac1r, \frac{c}{c-1}\} \xi_A({\bf x}) +O(1)$,
where ${\bf x} = \sum_{i=1}^{n}w_A(R_i)$.
Consider $D$ first,
\[
  \frac{D}{1-\epsilon} ={\bf b_0} \cdot \sum_{x \in I_{k+1}} y {\bf w}_A(x).
\]
Given a close slip $P$ with width $x$,
 i.e, the packed height in it is at least $c-1$,
\[
  \sum_{R \in P}{\bf w}_A(R) \ge (c-1){\bf w}_A(x).  
\]

Let $l_j$ be the number of type j slips with packed heights at least c-1 and 
$l_m = \sum_{j=0}^{k} l_j $ be the total number of these slips.
Let $x_h$ be the width of the $h$-th slip. 
Then
\begin{eqnarray*}
 {\bf b_i} \cdot \sum_{j=1}^{n} {\bf w}_A(R_j)
 &\ge& {\bf b_i} \cdot \sum_{h=1}^{l_m} (c-1){\bf w}_A(x_h) \\
&=& {\bf b_i} \cdot \sum_{x_h \in I_j, \phi(j) =i} (c-1){\bf w}_A(x_h) \\
&=& (c-1) \sum_{ 0 \le j \le k , \phi(j) =i } \frac{(1-\alpha_j)l_j}{\beta_j} 
\end{eqnarray*}

Consider the cost for packing  blue slips, say $ c  B_i$,
 by lemma \ref{lemma:openbins},
\[
   c B_i  = c \sum_{ 0 \le j \le k,
                               \phi(j) = i} \frac{(1-\alpha_j)l_j}{\beta_j} 
 +O(1) \le \frac{c}{c-1} \times  {\bf b_i} \cdot \sum_{j=1}^{n} {\bf w}_A(R_j) +O(1)
\]
So, in the same way, we have
\[
  c R_i  = c \sum_{ 1 \le j \le k,
                               \varphi(j) = i} \frac{\alpha_jl_j}{\beta_j} 
 +O(1) \le \frac{c}{c-1} \times {\bf r_i} \cdot \sum_{j=1}^{n} {\bf w}_A(R_j) +O(1)
\]
Hence, we have $ cost_A(L) \le \max\{\frac{c}{c-1},\frac1r\}
 \xi_A \Big ( \sum_{i=1}^n {\bf w}_A(R_i)\Big ) + O(1)$. 
 \end{proof}

\begin{thebibliography}{99}

\bibitem{B85}
 B.S. Baker, A new proof for the first-fit decreasing bin-packing
 algorithm. {\em J. Algorithms} 6, 49-70, 1985.

\bibitem{BBK81}
B.S. Baker, D.J. Brown, and H.P. Katseff, A 5/4  algorithm for
two-dimensional packing. {\em J. Algorithms} 2, 348-368, 1981.

\bibitem{BS83}
B.S. Baker and J.S. Schwarz, Shelf algorithms for two-dimensional
packing problems, {\em SIAM J. Comput.} 12, 508-525, 1983.

\bibitem{CGJT80}
E.G. Coffman, M.R. Garey, D.S. Johnson, and R.E. Tarjan,
Performance bounds for level oriented two dimensional packing
algorithms, {\em SIAM J. Comput.} 9, 808-826, 1980.

\bibitem{cw97}
J. Csirik and G.J. Woeginger, Shelf algorithm for on-line strip
packing, {\em Information Processing Letters} 63, 171-175, 1997.

\bibitem{Golan81}
I. Golan, Performance bounds for orthogonal, oriented
two-dimensional packing algorithms, {\em SIAM J. Comput.} 10,
571-582, 1981.

\bibitem{J73}
D.S. Johnson, Near-optimal bin-packing algorithms, {\em doctoral
thesis, M.I.T., Cambridge, Mass.}, 1973.

\bibitem{kk82}
N. Karmarkar and R.M. Karp, An efficient approximation scheme for
the one-dimensional bin-packing problem, In {\em Proc. 23rd Annual
IEEE Symp. Found. Comput. Sci.}, 312-320, 1982.

\bibitem{KR00}
C. Kenyon and E.R\'emila, A near-optimal solution to a
two-dimensional cutting stock problem, {\em Mathematics of
Operations Research} 25, 645-656, 2000.

\bibitem{LL85}
C.C. Lee and D.T. Lee, A simple on-line bin-packing algorihtm,
{\em J. ACM} 32, 562-572, 1985.

\bibitem{RBLL89}
P.V. Ramanan, D.J. Brown, C.C. Lee, and D. T. Lee, On-line bin
packing in linear Time, {\em J. Algorithms} 10, 305-326, 1989.

\bibitem{R91} M.B. Richey, Improved bounds for harmonic-based bin packing
algorithms, {\em Discrete Appl. Math.} 34, 203-227, 1991.

\bibitem{s02}
S.S. Seiden, On the online bin packing problem, {\em J. ACM} 49,
640-671, 2002.


%% \bibitem{SW05}
%% S.S. Seiden and G.J. Woeginger, The two-dimensional cutting stock
%% problem revisited, {\em Math. Program.} Ser.A 102, 519-530, 2005.

\bibitem{ullman71}
J.D. Ullman, The performance of a memory allocation algorithm. {\em
Tech. Rep. 100, Princeton University, Princeton, N.J.,Oct.}, 1971.

\bibitem{Vliet92}
A. van Vliet, An improved lower  bound for on-line bin packing
algorithms, {\em Inform. Process. Lett.} 43, 277-284,1992.

\bibitem{Yao80} A.C.-C. Yao, New Algorithms for Bin Packing,
{\em J. ACM} 27, 207-227, 1980.

\bibitem{Y91}
M. Yue, A simple proof of the inequality FFD(L) $\le$ 11/9OPT(L)
+1, $\forall L$ for the FFD bin-packing algorithm, {\em Acta
mathematicae applicatae sinica} 7, 321-331, 1991.

\end{thebibliography}
\end{document}